\documentclass[11pt]{amsart}
\reversemarginpar
\newcommand{\commentout}[1]{}

\usepackage{amsmath}
\usepackage{amssymb}

\newcommand{\nwc}{\newcommand}

\newcommand{\lt}{\left}
\nwc{\partz}{\frac{\partial }{\partial z}}
\newcommand{\rt}{\right}
\nwc{\ytil}{\tilde{\by}}
\nwc{\al}{\alpha}
\nwc{\half}{\frac{1}{2}}

\newcommand{\kvec}{\vec{\bk}}

\newcommand{\ks}{{k}}
\newcommand{\bx}{\mathbf x}
\newcommand{\bD}{\mathbf D}
\newcommand{\bp}{\mathbf p}

\newcommand{\bq}{\mathbf q}
\newcommand{\by}{\mathbf y}

\nwc{\nwt}{\newtheorem}

\nwt{proposition}{Proposition}
\nwt{lemma}{Lemma}
\nwt{theorem}{Theorem}

\nwt{remark}{Remark}
\nwt{assumption}{Assumption}
\nwt{definition}{Definition} 

\nwc{\bal}{\begin{align}}
\nwc{\be}{\begin{equation}}
\nwc{\ben}{\begin{equation*}}
\nwc{\bea}{\begin{eqnarray}}
\nwc{\beq}{\begin{eqnarray}}
\nwc{\bean}{\begin{eqnarray*}}
\nwc{\beqn}{\begin{eqnarray*}}
\nwc{\beqast}{\begin{eqnarray*}}

\nwc{\eal}{\end{align}}
\nwc{\ee}{\end{equation}}
\nwc{\een}{\end{equation*}}
\nwc{\eea}{\end{eqnarray}}
\nwc{\eeq}{\end{eqnarray}}
\nwc{\eean}{\end{eqnarray*}}
\nwc{\eeqn}{\end{eqnarray*}}
\nwc{\eeqast}{\end{eqnarray*}}

\nwc{\invf}{\cF^{-1}_2}
\nwc{\ep}{\varepsilon}
\nwc{\tep}{\tilde{\varepsilon}}
\nwc{\epsq}{{\varepsilon^2}}
\nwc{\epsqa}{{\varepsilon^{2\alpha}}}
\nwc{\eps}{\varepsilon}
\nwc{\ept}{\epsilon}
\nwc{\vrho}{\varrho}
\nwc{\orho}{\bar\varrho}
\nwc{\ou}{\bar u}
\nwc{\vpsi}{\varpsi}
\nwc{\lamb}{\lambda}

\nwc{\nn}{\nonumber}
\nwc{\bm}{\boldmath}
\nwc{\mf}{\mathbf}
\nwc{\mb}{\mathbf}
\nwc{\ml}{\mathcal}

\nwc{\IA}{\mathbb{A}} 
\nwc{\IB}{\mathbb{B}}
\nwc{\IC}{\mathbb{C}} 
\nwc{\ID}{\mathbb{D}} 
\nwc{\IM}{\mathbb{M}} 
\nwc{\IP}{\mathbb{P}} 
\nwc{\II}{\mathbb{I}} 
\nwc{\IE}{\mathbb{E}} 
\nwc{\IF}{\mathbb{F}} 
\nwc{\IG}{\mathbb{G}} 
\nwc{\IN}{\mathbb{N}} 
\nwc{\IQ}{\mathbb{Q}} 
\nwc{\IR}{\mathbb{R}} 
\nwc{\IT}{\mathbb{T}} 
\nwc{\IZ}{\mathbb{Z}} 

\nwc{\epal}{\ep^{-2\alpha}}
\nwc{\cE}{{\ml E}}
\nwc{\cP}{{\ml P}}
\nwc{\cQ}{{\ml Q}}
\nwc{\cL}{{\ml L}}
\nwc{\cR}{{\ml R}}
\nwc{\cV}{{\ml V}}
\nwc{\cT}{{\ml T}}
\nwc{\crV}{{\ml L}_{(\delta,\rho)}}
\nwc{\cC}{{\ml C}}
\nwc{\cA}{{\ml A}}
\nwc{\cK}{{\ml K}}
\nwc{\cB}{{\ml B}}
\nwc{\cD}{{\ml D}}
\nwc{\cF}{{\ml F}}
\nwc{\cS}{{\ml S}}
\nwc{\cM}{{\ml M}}
\nwc{\cG}{{\ml G}}
\nwc{\cH}{{\ml H}}
\nwc{\bk}{{\mb k}}
\nwc{\cbz}{\overline{\cB}_z}

\nwc{\pft}{\cF^{-1}_\bp}

\begin{document}

%
\title{Radiative Transfer  Limits of Two-Frequency
Wigner Distribution  for Random Parabolic Waves}

\author{Albert C. Fannjiang
 }
\thanks{Department of Mathematics,
University of California,
Davis, CA 95616
Email: fannjiang@math.ucdavis.edu.
The research is supported in part by National Science Foundation grant no. DMS-0306659, ONR Grant N00014-02-1-0090
and  Darpa Grant 
 N00014-02-1-0603
}

\begin{abstract}
The present note establishes the self-averaging, radiative transfer limit
for the two-frequency Wigner distribution for classical
waves in random media. 
Depending on the ratio of the wavelength to the correlation length
the limiting equation is either a Boltzmann-like 
integral equation or a Fokker-Planck-like differential
equation in the phase space. The limiting equation
is used to estimate
three physical parameters: the spatial spread,
the coherence length and the coherence bandwidth. 
In the longitudinal case, the Fokker-Planck-like equation can be solved
exactly. 

Dans cette note nous \'etablissons la limite auto-moyennisante dans
le regime du transfert radiatif pour la distribution de Wigner a deux
fr\'equences dans le cas classique d'ondes en milieux al\'eatoires.
Suivant le rapport de la longueur d'onde \`a la longueur de
corr\'elation l'\'equation limite est soit une \'equation int\'egrale
de type Boltzmann soit une \'equation diff\'erentielle dans l'espace
d'\'etat du type Fokker-Planck. L'\'equation limite est utilis\'ee
pour estimer trois param\`etres physiques: l'\'etalement spatial, la
longueur de coh\'erence et la largeur de bande coh\'erente. Dans le
cas longitudinal l'\'equation de type Fokker-Planck admet une solution
exacte.
\end{abstract}

\maketitle

\section{Introduction}
High-data-rate communication systems at millimeter and
optical frequencies, remote sensing and  detection
and the astronomical imaging all require
understanding of stochastic pulse propagation.
As pulses consist of
a typically broad frequency band
the complete information about transient
propagation requires a solution for the
statistical moments of the wave field at different
frequencies and locations \cite{Ish}.

Let  $k_1,k_2$ be two 
(relative) 
 wavenumbers  nondimensionlized by
the central wavenumber $k_0$.   
Let the wave fields $\Psi_j, j=1,2,$ of
$k_j, j=1,2,$
satisfy the paraxial wave equation 
in the dimensionless form \cite{Ish}
\beq
\label{para2}
i\partz
\Psi_j(z,\bx)+\frac{\gamma}{2\ks_j}\nabla^2\Psi_j(z,\bx)
+
\frac{\mu k_j}{\gamma}V(\frac{z}{\delta}, \frac{\bx}{\ep})\Psi_j(z,\bx)=0, \quad j=1,2
\eeq
where $\gamma$ is the Fresnel number,
$V$ represents 
 the refractive index  fluctuation with
  the correlation lengths $\delta$ and $\ep$
  in the longitudinal and transverse direction,
  respectively, and $\mu$ is the magnitude. 
Both  $\delta$ and $\ep$ are small parameters
related by the anisotropy parameter $\alpha$ as 
$\delta\alpha=\ep$. When $\alpha\ll 1$ (resp. $\alpha\gg 1$), the
refractive index fluctuates much faster (resp. slower) in the transverse
direction(s)  than in the longitudinal direction.

An important regime for classical wave propagation takes
place when the correlation length  is much smaller than
the propagation distance but is comparable or
much larger than the central wavelength
which is proportional to the Fresnel number.
This is the 
radiative transfer regime  for monochromatic waves described by 
the following scaling limit 
\beq
\gamma=\theta\ep,\quad \mu=\sqrt{\delta}, \quad\theta>0
\label{scale}
,\quad \hbox{such that}\quad  \lim_{\ep\to 0} \theta<\infty,
\eeq
with suitably chosen $\mu$ (see \cite{rad-arma}, \cite{rad-crm},
 \cite{Mis}, \cite{RPK} and references therein).  With two different frequencies, 
the most interesting scaling limit requires 
another simultaneous limit
\beq
\label{band}
\lim_{\ep\to 0}\ks_1=\lim_{\ep\to 0}\ks_2=\ks,
\quad \lim_{\ep\to 0}\gamma^{-1}(\ks_2-\ks_1)=\beta>0. 
\eeq
We shall refer to the conditions (\ref{scale}) and (\ref{band})
as the two-frequency radiative transfer scaling  limit.

The two-frequency
mutual coherence function \cite{Ish}
\[
\Gamma_{12}(z,\bx,\by)=
\IE[\Psi_1(z,\bx+\frac{\gamma\by}{2})\Psi_2(z,\bx
-\frac{\gamma\by}{2})],
\]
where $\IE$ stands for the ensemble averaging, plays an important role in analyzing propagation of
random pulses \cite{Ish}. But in the radiative transfer regime
the two-frequency
mutual coherence function is not as
convenient  as the two-frequency Wigner
distribution,  introduced in \cite{2f-whn}, which
is a natural extension of
the standard Wigner distribution and
is
self-averaging in the radiative transfer regime.  

\subsection{Two-frequency Wigner distribution}
The two-frequency Wigner distribution is
defined as  
\beq
\label{0.11}
W(z, \bx,\bp)=\frac{1}{(2\pi)^d}\int
e^{-i\bp\cdot\by}
\Psi_1 (z,\frac{\bx}{\sqrt{k_1}}+
\frac{\gamma\by}{2\sqrt{k_1}}){\Psi_2^{*}(z,\frac{\bx}{
\sqrt{k_2}}
-\frac{\gamma\by}{2\sqrt{k_2}})}d\by
\eeq
where, as well as below,  we have adopted  the easier notation $\xi_j=({k_j})^{-1/2}$. The choice of the scaling factors
in the definition is crucial.

The Wigner distribution 
has the following  properties:
\bea\label{2.2.2}
\int W_z(\bx,\bp)e^{i\bp\cdot\by}d\bp&=&
\Psi_1
(z,\frac{\bx}{\sqrt{k_1}}+
\frac{\gamma\by}{2\sqrt{k_1}})
\Psi_2^{*}(z,\frac{\bx}{\sqrt{k_2}}-\frac{\gamma\by}{2\sqrt{k_2}})\\
\int_{\IR^d}W_z(\bx,\bp)e^{-i\bx\cdot
\bq}d\bx&=&\lt(\frac{\sqrt{k_1k_2}\pi^2}{\gamma}\rt)^{d}
\widehat\Psi_1(z,\frac{\sqrt{k_1}\bp}{4\gamma}
+\frac{\sqrt{k_1}\bq}{2})
{\widehat\Psi}^{*}_2(z,\frac{\sqrt{k_2}\bp}{4\gamma}
-\frac{\sqrt{k_2}\bq}{2})
\eeq
and hence contains all the information 
in the two-point two-frequency function. 
Furthermore, the two-frequency Wigner distribution $W$
satisfies the two-frequency Wigner-Moyal equation
exactly  \cite{2f-whn}
\beq
\frac{\partial W}{\partial z}
+{\bp}\cdot\nabla_\bx W
+\frac{1}{\sqrt{\alpha\ep}}\cV_z W=0
\label{wig}
\eeq
where the operator $\cV_z$ is  given as
\beqn
\cV_z W
&=&
i\int 
\theta^{-1}\lt[e^{i\bq\cdot\bx/(\sqrt{k_1}\ep)}\ks_1
W(\bx,\bp+\frac{\theta\bq}{2\sqrt{k_1}})-
e^{i\bq\cdot\bx/(\sqrt{k_2}\ep)}\ks_2 W(\bx,\bp-
\frac{\theta\bq}{2\sqrt{k_2}})\rt]
\widehat{V}(\frac{\alpha z}{\ep},d\bq).
\eeqn

\section{Assumptions on the refractive index fluctuation}
We assume that $V_z(\bx)=V(z,\bx)$ is a centered,  $z$-stationary,
$\bx$-homogeneous random field
admitting the spectral representation
\[
V_z(\bx)=\int \exp{(i\bp\cdot\bx)}\hat{V}_z(d
\bp)
\]
with the  $z$-stationary spectral measure
$\hat{V}_z(\cdot)$ satisfying
\[
\IE[\hat{V}_z(d\bp)\hat{V}_z(d\bq)]
=\delta(\bp+\bq)\Phi_0(\bp)d\bp d\bq.
\]

The transverse power spectrum  density is
related to the full power spectrum density
$\Phi(\xi,\bp)$ as $\Phi_0(\bp)=\int \Phi(w,\bp)dw.
$
The power spectral density $\Phi(\kvec)$
satisfies $\Phi(\kvec)=\Phi(-\kvec),\forall
 \kvec=(w, \bp)\in \IR^{d+1}$
because
 the electric susceptibility field is assumed to be real-valued.
Hence
$
\Phi(w,\bp)=\Phi(-w,\bp)=\Phi(w,-\bp)
=\Phi(-w,-\bp)
$
which is related to the detailed balance of
the limiting scattering operators
described below.

More specifically we make the following two assumptions.
\begin{assumption}
 $V(z, \bx) $ is a   Gaussian process  with a
spectral density $\Phi(\kvec), \kvec=(w,\bp)\in\IR^{d+1}$ which is uniformly bounded
and decays at $|\kvec|=\infty$ with sufficiently
high power of $|\kvec|^{-1}$.
\end{assumption}
We note that the assumption of Gaussianity is not
essential and is made here to simplify the presentation.

Let $\cF_z $ and $\cF^+_z$  be the sigma-algebras generated by
$\{V_s:  \forall s\leq z\}$ and $\{V_s: \forall s\geq  z\}$,
respectively and let $L^2(\cF_z)$ and $L^2(\cF^+_z)$
denote the square-integrable functions measurable w.r.t.
to them respectively.
The maximal  correlation coefficient $r(t)$ is given by
\beq
\label{correl}
r(t)=\sup_{h\in L^2(\cF_z)\atop \IE[h]=0,
\IE[h^2]=1}
\sup_{
g\in L^2(\cF_{z+t}^+)\atop \IE[g]=0,
\IE[g^2]=1}\IE\lt[h g\rt].
\eeq

\begin{assumption}
\label{ass1}

The maximal correlation coefficient
$r(t)$  is integrable:
$\int^\infty_0r(s)ds
<\infty.
$
\end{assumption}

\section{Main theorems}

\begin{theorem}
Let $\theta>0$ be fixed.    Then  under
the two-frequency radiative transfer scaling (\ref{scale})-(\ref{band}) the 
weak solutions of the Wigner-Moyal equation 
(\ref{wig})
converges in law in the space
$C([0,\infty),L^2_w(\IR^d))$  to that of
the following deterministic equation
\beq
\label{bolt}
{\partz W+\bp\cdot\nabla  W}
&=&
\frac{2\pi k^2}{\theta^2}\int
K(\bp,\bq)
\lt[
e^{-i\beta\theta\bq\cdot\bx/(2k^{3/2})}
W(\bx,\bp+\frac{\theta \bq}{\sqrt{k}})
-W(\bx,\bp)
\rt]
{d\bq}
\eeq
where the kernel $K$ is given by
\beq
K(\bp,\bq)&=&\frac{1}{\alpha}\Phi\big(\alpha^{-1}(\bp+\frac{\theta\bq}{2\sqrt{k}})\cdot\bq,\bq\big).
\label{10'}
\eeq
\end{theorem}
 Here and below $L^2_w(\IR^{2d})$ is the space of
complex-valued square integrable
functions on the phase space $\IR^{2d}$ endowed with
the weak topology.

\begin{remark}
If we now let $\alpha\to 0$, then the kernel becomes
\beq
\label{11'}
K(\bp,\bq)&=&
\delta\big((\bp+\frac{\theta\bq}{2\sqrt{k}})\cdot\bq\big)
\int\Phi(w,\bq)
dw.
\nn
\eeq
We refer to this as the transverse case
because the transverse correlation length $\ep$
is much shorter than the longitudinal  correlation length
$\delta$.

On the other hand, in the longitudinal case
$\alpha\to \infty$ the limiting kernel  would vanish. 
In order to maintain an interesting limit, we
increase $\mu$ by a factor of $\sqrt{\alpha}$.
Then the kernel for the longitudinal case becomes
\beq
\label{B0}
K(\bp,\bq)&=&
\Phi(0,\bq).
\eeq

In both the longitudinal and transverse cases
the fluctuations in
the refractive index are  extremely anisotropic. 

\end{remark}
\begin{theorem}
Assume $\lim_{\ep\to 0} \theta=0$.  Then under the two-frequency radiative transfer
scaling (\ref{scale})-(\ref{band})  the 
weak solutions of the Wigner-Moyal equation 
(\ref{wig})
converges in law in the space
$C([0,\infty),L^2_w(\IR^d))$ to that of the following
deterministic equation
\beq
\label{fp}
\partz W+\bp\cdot\nabla W=-\ks\lt(i\nabla_\bp-\frac{\beta}{2\ks}\bx\rt)
\cdot \bD\cdot 
\lt(i\nabla_\bp-\frac{\beta}{2\ks}\bx\rt) W
(\bx,\bp)
\eeq
where the diffusion coefficient $\bD$ is
given by
\beq
\label{D3}
\bD(\bp)&=&\frac{\pi}{\alpha}\int \Phi(\frac{\bp\cdot\bq}{\alpha},\bq)
\bq\otimes\bq d\bq,.
\eeq
\end{theorem}

\begin{remark}
In the transverse case $\alpha\to 0$,
the limiting coefficient is 
 \beq
\label{D2}
\bD(\bp)&=&\pi |\bp|^{-1}
\int_{\bp\cdot\bp_\perp=0} \int
\Phi(w,\bp_\perp) d w\,\,
\bp_\perp\otimes\bp_\perp
d\bp_\perp.
\eeq

For the longitudinal case $\alpha\to\infty$,
we increase $\mu$ 
by a factor of $\sqrt{\alpha}$ as before such that
the limiting coefficient is  nontrivial 
\beq
\label{D1}\bD&=\pi\int \Phi(0,\bq)\bq\otimes\bq d\bq.
\eeq

\end{remark}

 When $k_1=k_2$ or $\beta=0$, eq. (\ref{bolt}) and (\ref{fp})
reduce to the standard radiative transfer equations derived
in \cite{rad-arma}, \cite{rad-crm}. The proof of these results follows exactly the strategy 
developed in \cite{rad-arma} and  outlined
in \cite{rad-crm}, originally developed for
the standard one-frequency Wigner distribution
(see \cite{2f-whn} for the same strategy applied
to the two-frequency Wigner distribution for
a different scaling limit). 

Another notable fact is that eq. (\ref{bolt}) with (\ref{B0})
and eq. (\ref{fp}) with (\ref{D1})
are similar to the  governing equations for
the {\em ensemble-averaged }
two-frequency Wigner distribution for
the $z$-white-noise potential analyzed in \cite{2f-whn}. 
This can be understood by the similar behaviors
of the potential more rapidly fluctuating in $z$ to the
$z$-white-noise potential and the (less) rapid fluctuation
in $\bx$ gives rise to self-averaging which is lacking
in the $z$-white-noise potential.

\section{The longitudinal and transverse case}
To illustrate the utility
of these equations, we proceed to discuss
the two special cases in three dimensions. For simplicity, we will assume the isotropy of the medium in the transverse coordinates such that $\Phi(w,\bp)=\Phi(w,|\bp|)$.
As a consequence $\bD=D {\mathbf I}$ with a constant scalar $D$ in the longitudinal case   whereas 
in  the transverse case $\bD(\bp)= C|\bp|^{-1}\hat\bp_\perp\otimes \hat\bp_\perp$
with the constant  $C$ given by
\[
C=\frac{\pi}{2}\int\int \Phi(w, \bp_\perp)dw |\bp_\perp|^2 d\bp_\perp.
\]
Here $\hat\bp_\perp\in \IR^2$ is an unit vector normal to $\bp\in\IR^2$.

First of all,  the equation (\ref{fp}) by itself
gives qualitative information about three 
important parameters of the stochastic channel:
the spatial spread $\sigma_*$, the coherence length $\ell_c$
and the coherence bandwidth $\beta_c$, through the
following scaling argument. One seeks the change of
variables
\beq
\label{new}
\tilde\bx=\frac{\bx}{\sigma_*\sqrt{k}},\quad
\tilde\bp=\bp \ell_c\sqrt{k},  \quad \tilde z=\frac{z}{L}, \quad
\tilde\beta=\frac{\beta}{\beta_c}\eeq
where $L$ is the propagation distance
to remove all the physical parameters from
(\ref{fp}) and to aim for
the form 
\beq
\label{fp'}
\frac{\partial }{\partial \tilde z} W+\tilde\bp\cdot\nabla_{\tilde\bx} W
=-\lt(-i\nabla_{\tilde\bp}+\frac{\tilde\beta}{2}\tilde\bx\rt)\cdot
\lt(-i\nabla_{\tilde\bp}+\frac{\tilde\beta}{2}\tilde \bx\rt) W
\eeq
 in the longitudinal case
 and the form
 \beq
\label{fpn}
\frac{\partial }{\partial \tilde z} W+\tilde\bp\cdot\nabla_{\tilde\bx} W
=-\lt(-i\nabla_{\tilde\bp}+\frac{\tilde\beta}{2}\tilde\bx\rt)\cdot
\frac{\hat\bp_\perp\otimes\hat\bp_\perp}{|\tilde\bp|}\cdot
\lt(-i\nabla_{\tilde\bp}+\frac{\tilde\beta}{2}\tilde \bx\rt) W
\eeq
 in the transverse case.
 From the left side of (\ref{fp})
it immediately follows the first duality relation $\ell_c\sigma_*\sim L/k$.
The balance of terms inside each pair of parentheses leads
to the second duality relation $\beta_c\sim \ell_c k/\sigma_*$.
Finally the removal of $D$ or $C$ determines the spatial spread $\sigma_*$ which has a different expression in
the longitudinal and transverse case.  
In the longitudinal case, $\sigma_*\sim \sqrt{DL^3}$
whereas in the transverse case $\sigma_*\sim 
C^{1/3}L^{4/3}k^{-1/6}$.

In the longitudinal case, the inverse Fourier transform in $\tilde\bp$ renders  eq. (\ref{fp'})  to the form
\beq
\label{mean-eq2}
{\frac{\partial \tilde W}{\partial \tilde z}
-{i}\nabla_{\tilde\by}\cdot\nabla_{\tilde\bx} \tilde
W}
&=&-\big|\tilde\by+\frac{\tilde\beta}{2}\tilde\bx\big|^2
 \tilde W
\eeq
which can be solved exactly and whose Green function is \cite{2f-whn}
\beq
&&\frac{(1+i)^{d/2}\tilde\beta^{d/4}}{(2\pi)^d\sin^{d/2}{\big[\tilde\beta^{1/2}(1+i)\big]}}  \nn e^{i\frac{
|\tilde\by-\by'|^2}{2\tilde\beta}}
e^{i\frac{
(\tilde\by-\by')\cdot(\tilde\bx-\bx')}{2}}
e^{i\frac{\tilde{\beta}|\tilde\bx-\bx'|^2}{8}}
\\
\nn&&\times e^{\frac{1-i}{2\sqrt{\tilde\beta}}
\cot{(\sqrt{\tilde\beta}(1+i))}\big|
\tilde\by-\tilde\beta\tilde\bx/2
-\frac{\by'-\tilde\beta\bx'/2}{
\cos{(\sqrt{\tilde\beta}(1+i))}}\big|^2}
 e^{-\frac{1-i}{2\sqrt{\tilde\beta}}
\lt|\by'-\tilde\beta\bx'/2\rt|^2
\tan{(\sqrt{\tilde\beta
}(1+i))}}\nn
\eeq
for $\tilde z=1$. 
This solution gives asymptotically 
precise information about the cross-frequency
correlation, important for  analyzing
the  information transfer  and time reversal with
broadband signals in the channel 
described by the random Schr\"odinger equation
 \cite{pulsa} (see also \cite{DTF2}, \cite{DTF}, \cite{BPZ}). 
It is unclear if the transverse case is exactly solvable or not.

\end{document}